\documentclass[prc,superscriptaddress,twocolumn,showpacs,preprintnumbers,amsmath,amssymb,floatfix,dvipdfmx]{revtex4-2}
\usepackage{amsmath}
\usepackage{amssymb}
\usepackage{mathrsfs}
\usepackage{bm}
\usepackage{graphicx}
\usepackage{braket}
\usepackage{mathrsfs}
\usepackage{color}
\usepackage{ulem}

\makeatletter
\let\MYcaption\@makecaption
\makeatother
\usepackage{subcaption}
\makeatletter
\let\@makecaption\MYcaption
\makeatother

\def\mbold#1{\mbox{\boldmath $#1$}}

\makeatletter
\newcommand{\figcaption}[1]{\def\@captype{figure}\caption{#1}}
\newcommand{\tblcaption}[1]{\def\@captype{table}\caption{#1}}
\makeatother

\begin{document}

\title{
Dineutron in the $2^+_1$ state of $^6$He
}

\author{Shoya Ogawa}
\email[]{s-ogawa@phys.kyushu-u.ac.jp}
\affiliation{Department of Physics, Kyushu University, Fukuoka 819-0395, Japan}

\author{Takuma Matsumoto}
\email[]{matsumoto@phys.kyushu-u.ac.jp}
\affiliation{Department of Physics, Kyushu University, Fukuoka 819-0395, Japan}

\date{\today}

\begin{abstract}
 We investigate the dineutron in the $2^+_1$ state of $^6$He via analysis of its decay mode
 by using the complex scaling method.
 In this letter, we propose the cross section for the resonant state to distinguish
 the resonant contributions from the nonresonant ones.
 As the results, it is found that the shoulder peak appears
 in the cross section for the resonant state as a function of $\varepsilon_{n \text{-} n}$.
 Furthermore, we show that the $S=0$ component of the cross section,
 where $S$ is the total spin of the valence two neutrons, has a peak around the shoulder peak,
 which comes from the dineutron configuration in the $2^+_1$ state.
 Thus we conclude that the shoulder peak is expected to indicate the existence of 
 the dineutron in the $2^+_1$ state.
\end{abstract}

\maketitle

%
{\it Introduction.}
Neutron-rich nuclei have been intensively pursued since the development
of radioactive ion-beam experiments. 
Two-neutron halo nuclei appear near the neutron dripline and have
loosely bound two neutrons surrounding a core nucleus. 
As properties of two-neutron halo nuclei, the structure is described 
by a $n$ + $n$ + core three-body system and is referred to as the Borromean 
structure, which has no bound subsystems. 
Besides, there is only one bound state, i.e., the ground state.
In the ground state of two-neutron halo nuclei, 
existence of the dineutron, which is a spatially compact two-neutron pair, 
has been predicted in various theoretical
calculations~\cite{Oga99,Zhu93,Ber91,Esb99,Hag05,Hag09,Jag20,Cas20,Pap11,Kik10,Kik16,Kub20,Sun21,Coo20}. 
Recently, it has been clarified that the dineutron develops 
in the surface region of $^{11}$Li by the experiment for the knockout 
reaction~\cite{Kub20}. 
Furthermore, 
experimental studies for Coulomb breakup reactions indicate the existence of 
the dineutron in the ground states of $^6$He~\cite{Sun21} and $^{19}$B~\cite{Coo20}. 

Excited states of two-neutron halo nuclei appear above the three-body 
threshold as resonant states. The resonant states are unbound states and decay into
three particles, namely, two neutrons and a core nucleus. 
Elucidation of some resonant states, 
e.g. the $2^+_1$ state in $^6$He~\cite{Kik13}
and unbound nuclei $^6$Be~\cite{Ois14,Cas18,Wan21}, $^{16}$Be~\cite{Cas18,Lov17}, 
and $^{26}$O~\cite{Hag14,Hag16},
have been attracted much attention and investigated via decay-particle measurements, 
which include information of the structure. 
However the decay observables, such as excitation energy spectra of
the cross section, contain not only the resonant contribution
but also contributions from the nonresonant states.
To investigate structural information of the resonant states, we need to
eliminate the nonresonant contributions from the cross section~\cite{Oga20}.
This point makes it difficult to clarify properties of the resonant states.

$^6$He is the lightest two-neutron halo nucleus and has been investigated intensively 
so far~\cite{Aum99,Bac02,Mat04,Rod05,Des06,Mor07,Bay09,Bro12,Aco11,For14,Des16,Sun21}.
In Ref.~\cite{Kik13}, the $2_1^+$ resonant state of $^6$He
was investigated via the $^6$He + $^{12}$C reaction at 240 MeV/nucleon~\cite{Aum99}.
In the previous work, the double-differential breakup cross section
(DDBUX) with respect to the two-neutron relative energy
($\varepsilon_{n\mbox{-}n}$) and the energy between the center-of-mass
(c.m.) of the two-neutron system and $\alpha$
($\varepsilon_{nn\mbox{-}\alpha}$) was calculated by combining the
continuum discretized coupled channels method (CDCC)~\cite{Yah12} with
the complex-scaled Lippmann-Schwinger equation (CSLS)~\cite{Kik10,Kik16}.
Furthermore, to extract the contribution from the resonant state,
they calculated the breakup cross section
as a function of $\varepsilon_{n\mbox{-}n}$,
$d\sigma/d\varepsilon_{n\mbox{-}n}$,
by gating the total excited energy of $^6$He
within the range of the energy of the $2^+_1$ state, where
the DDBUX was integrated over $\varepsilon_{nn\mbox{-}\alpha}$.
According to the
results, the shoulder peak appears in
$d\sigma/d\varepsilon_{n\mbox{-}n}$ 
around 0.8 MeV.
They suggested that the shoulder peak indicates the existence of
the dineutron in the $2^{+}_{1}$ state. 

Although the cross section gated within the resonant energy, it cannot 
completely exclude the nonresonant contributions from the cross section. 
Therefore, the evidence of the dineutron in the $2^+_1$ state is insufficient at this stage. 
To clarify this point, 
it is necessary to obtain isolate a resonant state in multi-channel systems and analyze
its contribution to the cross section.
In order to calculate the resonant states,
various approaches have been used so far, such as 
the complex scaling method (CSM)~\cite{Agu71,Bal71,Myo14} and 
methods based on the hyperspherical coordinate~\cite{Lov17,Des06,Pin16,Cas19}.
In this study,
we propose a method of extracting only the resonant contribution from the cross section
by using the CSM.
In the CSM, 
the resonant state can be completely separated from the nonresonant state. 
Therefore we can evaluate the cross section to the resonant state calculated by the CSM.

In this letter,
the dineutron in the $2^+_1$ state of $^6$He is investigated via the
analysis of the $^6$He + $^{12}$C reaction at 240 MeV/nucleon
in the framework combining the CDCC with the CSLS. 
The reaction is described as a $n$ + $n$ + $\alpha$ + $^{12}$C four-body system, 
and the $2^+_1$ state is obtained by the CSM. 
In this analysis, we calculate the DDBUX and 
$d\sigma/d\varepsilon_{n\mbox{-}n}$ 
for the resonant contribution and discuss the dineutron configuration in the $2_1^+$ state.

%
{\it Formalism.}
The $^6$He + $^{12}$C system is described as the four-body breakup reaction,
and the Schr\"odinger equation is written as
\begin{eqnarray}
 \left[
  K_{R}+U+h-E
 \right]
 |\Psi^{(+)}\rangle=0 
 ,
\end{eqnarray}
with
\begin{eqnarray}
 U=U_{n}+U_{n}+U_{\alpha}+V_{\rm C},
\end{eqnarray} 
where $\bm{R}$ represents the coordinate between the c.m. of $^6$He and $^{12}$C.
$K_{R}$ is the kinetic energy operator associated with $\bm{R}$,
and $h$ is the internal Hamiltonian of $^6$He.
$U_n$ ($U_{\alpha}$) describes the optical potential between
$n$ ($\alpha$) and $^{12}$C.
These potentials are obtained by the folding model with Melbourne
$g$ matrix~\cite{Amo00} in the same manner as used in Ref.~\cite{Oga19}.
$V_{\rm C}$ is the Coulomb potential between the c.m. of $^6$He and
$^{12}$C, that is, Coulomb breakup is neglected in this study.

The CDCC equation is constructed within the model space $\mathcal{P}$ 
as 
\begin{eqnarray}
 \mathcal{P}
 \left[
  K_{R}+U+h-E
 \right]
 \mathcal{P}
 |\Psi^{(+)}\rangle=0,
\end{eqnarray}
where $\mathcal{P}$ is defined by
\begin{eqnarray}
 \mathcal{P}=\sum_{n}\ket{\Phi_{n}}\bra{\Phi_{n}}.
\end{eqnarray}
A set of eigenstates $\{\Phi_{n}\}$ is obtained by diagonalizing $h$
with the Gaussian expansion method (GEM)~\cite{Hiy03} and includes 
the bound and discretized continuum states. 
In the CDCC, the transition matrix to the discretized state 
is represented as
\begin{eqnarray}
 T_{n}
  =
  \braket{\Phi_{n}\chi_{n}^{(-)}(\mbold{P}_n)|
  U-V_{\rm C}|\mathcal{P}\Psi^{(+)}},
\end{eqnarray}
where $\chi_{n}^{(-)} (\mbold{P}_n)$ is the Coulomb wave function with
the asymptotic relative momentum $\mbold{P}_n$ and satisfies
the incoming boundary condition. 
Using the smoothing procedure with the CSLS~\cite{Kik13}, 
the continuous transition matrix is calculated as
\begin{eqnarray}
 \label{eq:tmatrix}
 T_{\varepsilon}(\bm{k},\bm{K},\bm{P})
  &=&
  \sum_{n}
  f_{\varepsilon,n}(\bm{k},\bm{K})
  T_{n}
  ,
\end{eqnarray}
with the smoothing function defined as
\begin{eqnarray}
 f_{\varepsilon,n}(\bm{k},\bm{K})
  &=& \braket{\Phi^{(-)}_{\varepsilon}(\bm{k},\bm{K})|\Phi_{n}}.
\end{eqnarray}
Here $\Phi^{(-)}_{\varepsilon}$ is the three-body scattering wavefunction
of $^6$He with the internal energy $\varepsilon$ and satisfies the
incoming boundary condition.
The asymptotic relative momentum regarding $\bm{R}$ is represented by
$\bm{P}$, and  
the asymptotic internal momenta of $\bm{k}$ and $\bm{K}$ in $^6$He
satisfy the relation $\varepsilon=(\hbar^{2}k^{2})/(2\mu_{n \text{-} n})
+(\hbar^{2}K^{2})/(2\mu_{nn \text{-} \alpha})$, where
$\mu_{n \text{-} n}$ and $\mu_{nn \text{-} \alpha}$ are the reduced masses of 
the $n \text{-} n$ and $nn \text{-} \alpha$ systems, respectively.

To calculate $f_{\varepsilon,n}(\bm{k},\bm{K})$, we apply the
CSLS that describes the three-body scattering wavefunction with the
appropriate boundary condition:
\begin{eqnarray}
 \label{eq:csls}
  &&
  f_{\varepsilon,n}(\bm{k},\bm{K})
  =
  \bra{\phi(\bm{k},\bm{K})}\Phi_{n}\rangle
  \nonumber \\
  &&~~+
  \sum_{\nu}
  \braket{\phi(\bm{k},\bm{K})|{V}U^{-1}_{\theta}|\Phi^{\theta}_{\nu}}
  \frac{1}{\varepsilon-\varepsilon^{\theta}_{\nu}}
  \bra{\tilde{\Phi}^{\theta}_{\nu}}
  U_{\theta}|\Phi_{n}\rangle
  ,~~~~
\end{eqnarray}
where $\phi$ represents the plane wave for three-body scattering.
${V}$ is the sum of the interactions in $h$.
$U_{\theta}$ is the scaling transformation operator in the CSM.
The $\nu$th eigenstate with the eigenenergy
$\varepsilon^{\theta}_{\nu}$ calculated by the CSM is represented by
$\Phi^{\theta}_{\nu}$.
It should be noted that a set of eigenstates $\{\Phi_\nu^{\theta}\}$ forms a
complete set as
$\sum_{\nu}\ket{\Phi^{\theta}_{\nu}}
\bra{\tilde{\Phi}^{\theta}_{\nu}}= 1$, which
is referred to as an extended completeness
relation~\cite{Myo98,Gir03,Gir04}. Furthermore, combining  
$U^{-1}_{\theta}U_{\theta}=1$ with 
the extended completeness relation,
we obtain
$\sum_{\nu}U^{-1}_{\theta}\ket{\Phi^{\theta}_{\nu}}
\bra{\tilde{\Phi}^{\theta}_{\nu}}U_{\theta}= 1$.

Using Eq.~\eqref{eq:tmatrix}, the DDBUX with respect to 
$\varepsilon_{n \text{-} n}$ and 
$\varepsilon_{nn \text{-} \alpha}$ is calculated as
\begin{eqnarray}
 \label{eq:DDX}
 &&
  \frac{d^{2}\sigma}
  {d\varepsilon_{n \text{-} n}d\varepsilon_{nn \text{-} \alpha}}
  =
  \sum_{n} \sum_{n'} T^{\dag}_{n} T_{n'} 
  \nonumber \\
 &&~~\times
  \int d\bm{k} d\bm{K} d\bm{P} ~
  f^{\dag}_{\varepsilon,n}(\bm{k},\bm{K})
  f_{\varepsilon,n'}(\bm{k},\bm{K}) 
  \nonumber \\
 &&~~\times
  \delta
  \left(
   E_{\rm tot}-\frac{\hbar^2 \bm{P}^2}{2\mu_{R}}
   -\varepsilon_{n \text{-} n}-\varepsilon_{nn \text{-} \alpha}
  \right)
  \nonumber \\
 &&~~\times
  \delta
  \left(
   \varepsilon_{n \text{-} n}-\frac{\hbar^{2}\bm{k}^{2}}{2\mu_{n \text{-} n}}
  \right)
  \delta
  \left(
   \varepsilon_{nn \text{-} \alpha}-\frac{\hbar^{2}\bm{K}^{2}}{2\mu_{nn \text{-} \alpha}}
  \right)
  ,
\end{eqnarray}
where $E_{\rm tot}$ is the total energy of the reaction system,
and $\mu_{R}$ is the reduced mass of the $^6$He + $^{12}$C system.

%
To extract the resonant contribution from Eq.~\eqref{eq:DDX},
we consider the transition matrix to $\Phi_\nu^\theta$, which is separated 
into the resonant and nonresonant states.
Inserting 
$\sum_{\nu}U^{-1}_{\theta}\ket{\Phi^{\theta}_{\nu}}
\bra{\tilde{\Phi}^{\theta}_{\nu}}U_{\theta}= 1$ 
into Eq.~\eqref{eq:tmatrix},
the continuous transition matrix and its Hermitian conjugate
are rewritten as  
\begin{eqnarray}
 \label{eq:tmatrix-res}
 T_{\varepsilon}(\bm{k},\bm{K},\bm{P})
 &=&
  \sum_{\nu}
  f^{\theta}_{\varepsilon,\nu}(\bm{k},\bm{K})
  \tilde{T}^{\theta}_{\nu}
  ,
\end{eqnarray}
with 
\begin{eqnarray}
 \label{eq:tmat-cs}
&& \tilde{T}^{\theta}_{\nu}
  =
  \sum_{n}
  \braket{\tilde{\Phi}^{\theta}_{\nu}|U_{\theta}|\Phi_{n}}
  T_{n}
  ,~~
  f^{\theta}_{\varepsilon,\nu}
  = \braket{\Phi^{(-)}_{\varepsilon}|U^{-1}_{\theta}|\Phi^{\theta}_{\nu}}
  .
  \nonumber \\
\end{eqnarray}
In Eq.~\eqref{eq:tmat-cs}, the arguments of $\bm{k}$
and $\bm{K}$ are omitted for simplicity.
$T^{\theta}_{\nu}$,
which has the same definition in Ref.~\cite{Mat10},
can be interpreted as the transition matrix to $\Phi^{\theta}_{\nu}$.
Using Eq.~\eqref{eq:tmatrix-res}, Eq.~\eqref{eq:DDX} is rewritten as
the following summation for $\nu$,
\begin{eqnarray}
 \label{eq:DDX-2}
 &&
 \frac{d^{2}\sigma}
 {d\varepsilon_{n \text{-} n}d\varepsilon_{nn \text{-} \alpha}}
 =
 \sum_{\nu} \sum_{\nu'} T^{\theta\dag}_{\nu}  T^{\theta}_{\nu'}   
 \nonumber \\
 &&~~\times
 \int d\bm{k} d\bm{K} d\bm{P}~
 f^{\theta\dag}_{\varepsilon,\nu}(\bm{k},\bm{K})
 f^{\theta}_{\varepsilon,\nu'}(\bm{k},\bm{K})
 \delta_{\rm e.c.}
 , 
\end{eqnarray}
where $\delta_{\rm e.c.}$ represents a set of the three $\delta$-functions in 
Eq.~\eqref{eq:DDX}.
We confirm that the result of Eq.~\eqref{eq:DDX-2} is consistent with 
that of Eq.~\eqref{eq:DDX}.
In this study, we define the DDBUX for the resonant state as 
\begin{eqnarray}
 \label{eq:DDX-res}
  &&
 \frac{d^{2}\sigma_{\nu_{\rm R}}}
 {d\varepsilon_{n \text{-} n}d\varepsilon_{nn \text{-} \alpha}}
 \equiv
 T^{\theta\dag}_{\nu_{\rm R}} T^{\theta}_{\nu_{\rm R}}
 \nonumber \\
 &&\times
  \int d\bm{k} d\bm{K} d\bm{P} ~
  f^{\theta\dag}_{\varepsilon,\nu_{\rm R}}(\bm{k},\bm{K})
  f^{\theta}_{\varepsilon,\nu_{\rm R}}(\bm{k},\bm{K})
  \delta_{\rm e.c.},
\end{eqnarray}
where $\nu_{\rm R}$ represents the resonant state $2^+_1$ with the resonant
energy $\varepsilon_{r}$ and decay width $\Gamma$.
This cross section is referred as the {\it resonant cross section} in this letter.

In this study,
we apply the same internal Hamiltonian $h$ as used in Ref.~\cite{Oga20}.
As a model space for the total spin $I$ and the parity $\pi$ in $^6$He, 
we take $I^{\pi}$ = $0^+$, $1^-$ and $2^+$.
The particle exchange between valence neutrons and neutrons in $\alpha$ 
is treated with the orthogonality condition model~\cite{Sai69}.
In the GEM, we take the Gaussian range parameters $r_i$ ($i$=1,2,...,$N$)
that lie in geometric progression.
We adopt the same parameters in Ref.~\cite{Kik13} for $\Phi_{n}$.
For $\Phi_{\theta,\nu}$ in the CSLS and $\Phi^{\theta}_{\nu_{\rm R}}$, 
($N$, $r_1$, $r_N$) = (22, 0.1 fm, 75 fm) and (16, 0.1 fm, 25 fm) are taken, respectively.
As the result, we obtain the ground state energy $-0.972$ MeV 
and ($\varepsilon_r$, $\Gamma$) = ($0.823$ MeV, $0.121$ MeV)
for the $2^+_1$. The scaling angle $\theta$ is set to 12 deg.
The convergence of the calculated cross section
has been achieved within about 5\% fluctuation.

%
%
{\it Results and Discussions.}
\begin{figure}[tbp]
 \begin{center}
  \includegraphics[scale=0.75]{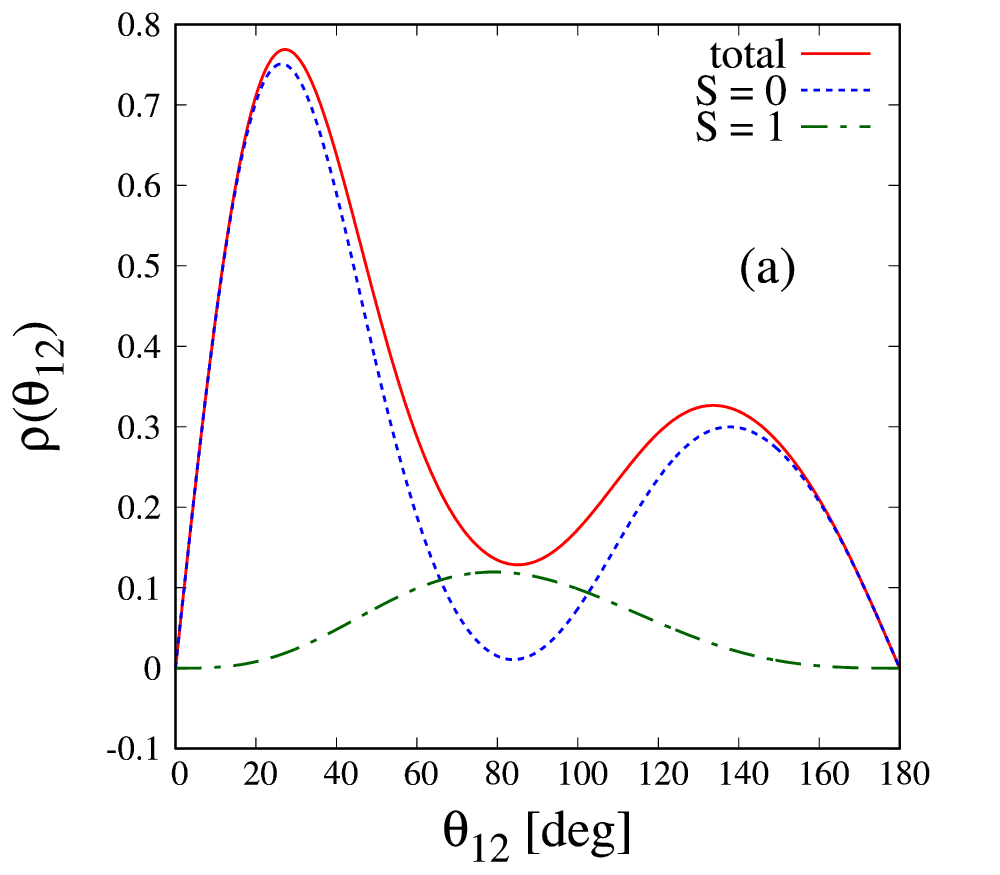}
  \includegraphics[scale=0.75]{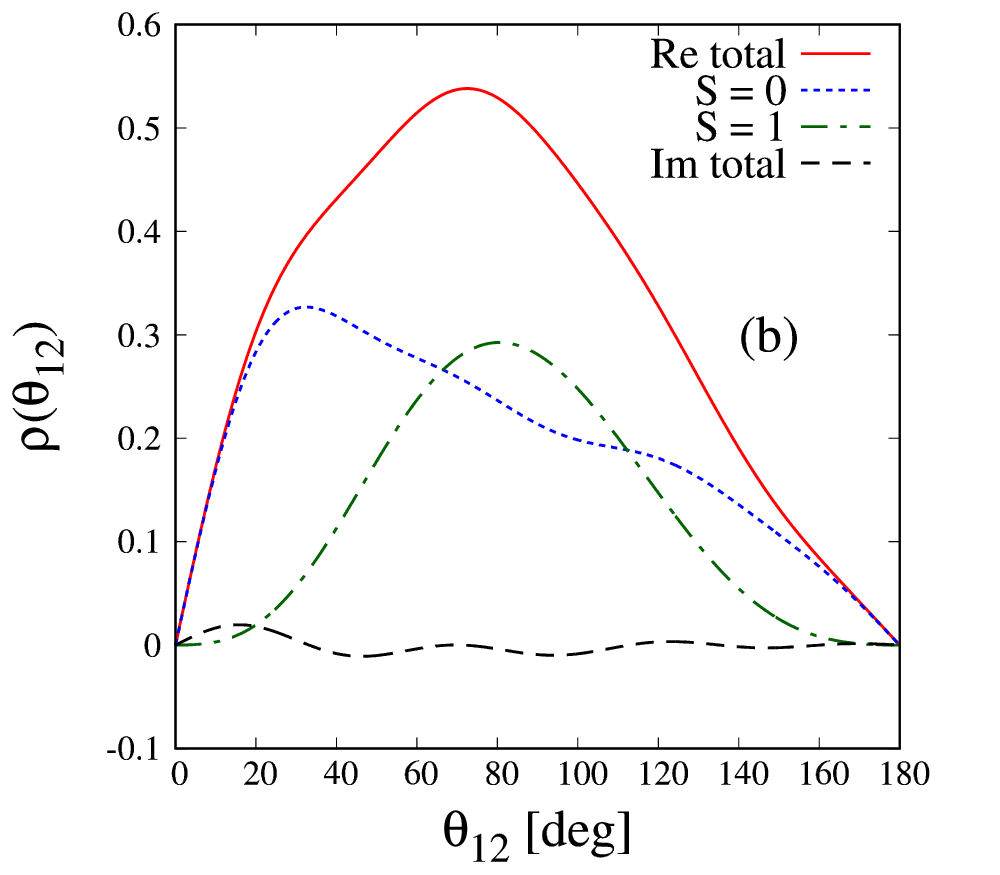}
  \caption{
  The angular density of (a) the ground state and (b) the $2^+_{1}$ state. 
  This density is a function as the opening angle between the two valence neutrons.
  }
  \label{fig:OpenAngle}
 \end{center}
\end{figure}
First, to discuss the dineutron in the $2^+_1$ state,
we consider the following angular density as
\begin{eqnarray}
 \label{eq:density}
 \rho(\theta_{12})
  &\equiv&\braket{\tilde{\Phi}^{\theta}_{\nu_{\rm R}}|
  \delta(\omega-\theta_{12})|\Phi^{\theta}_{\nu_{\rm R}}},
\end{eqnarray}
where $\theta_{12}$ is the opening angle between the two valence neutrons. 
This density is normalized as $\int\rho(\theta_{12})d\theta_{12}=1$
and independent of the scaling angle in the CSM.
The details of $\rho(\theta_{12})$ are discussed in Ref.~\cite{Kru14}.
Here it should be noted that the angular density of a resonant state is complex 
because an expected value for a resonant state is defined in the framework of 
Non-Hermitian Quantum Mechanics~\cite{Moi11}.
According to Ref.~\cite{Ber96}, 
the real part means the expected value of an operator,
and the imaginary part, 
which comes from the interference between the resonant state and nonresonant states, 
corresponds to the uncertainty of the expected value.

In  Fig.~\ref{fig:OpenAngle}(a),
we demonstrate the angular density of the ground state represented by
the solid line, which shows the two peaks at the small and large
angles. The peak at the small angle indicates the dineutron configuration 
because the small angle means the short distance between the valence two neutrons.
To discuss this behavior in more details, we separate the
angular density into the $S$ = 0 and 1 components, where $S$ represents
the total spin of the valence two neutrons.
The dotted and dot-dashed lines represent the angular density for $S$ = 0 and 1, 
respectively. 
One sees that the $S$ = 0 component has also the two
peaks at the small and large angles, and the $S$ = 1 component behaves
almost symmetrically. 
Therefore, the dineutron is formed in the case for $S=0$.

The solid line in Fig.~\ref{fig:OpenAngle}(b) represents the real part
of the angular density of the $2^+_1$ state, and it takes the maximum
value in the region $\theta_{12}$ = $60$--$80^{\circ}$.
Since the imaginary part of $\rho(\theta_{12})$ shown by the dashed line is
negligibly small, we discuss only the real part of $\rho(\theta_{12})$.
The dotted and dot-dashed lines represent the angular density for $S=0$
and 1, respectively. One can see that the $S=0$ component has a peak
structure at the small angle. Therefore the dineutron in the $2^+_1$ state is
expected to be clear when we focus on the $S=0$ component.

\begin{figure}[tbp]
 \begin{center}
  \includegraphics[scale=0.8]{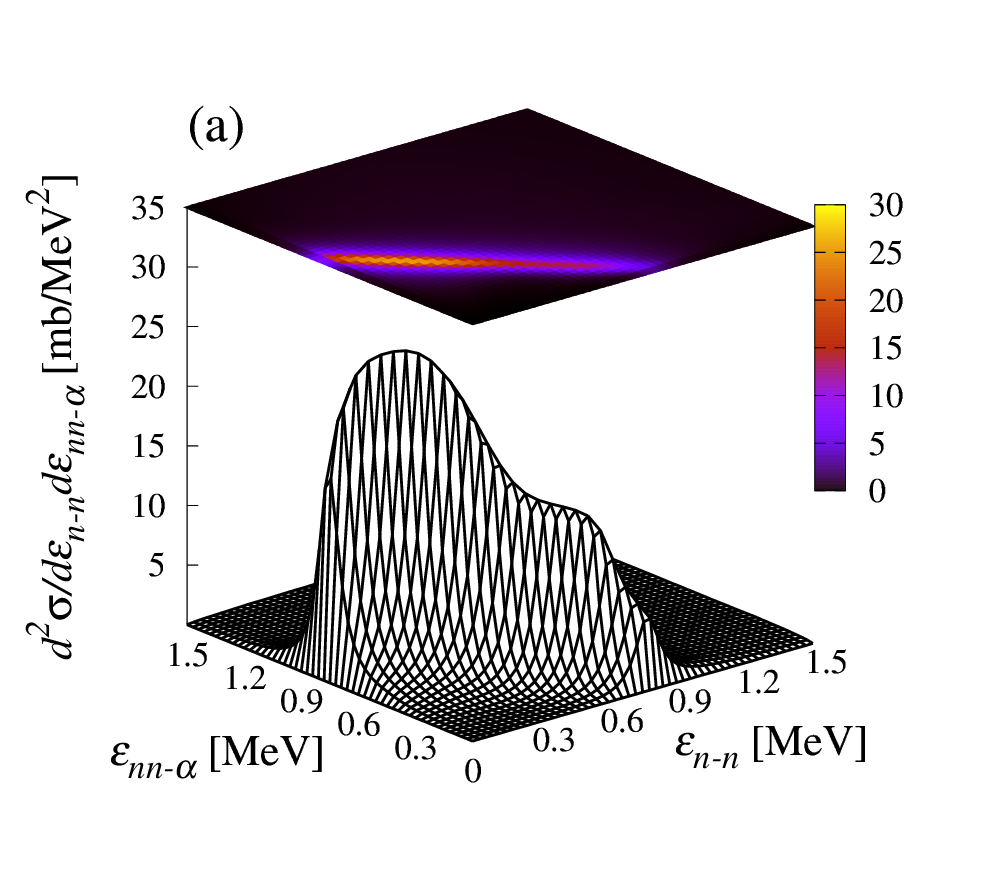}
  \includegraphics[scale=0.8]{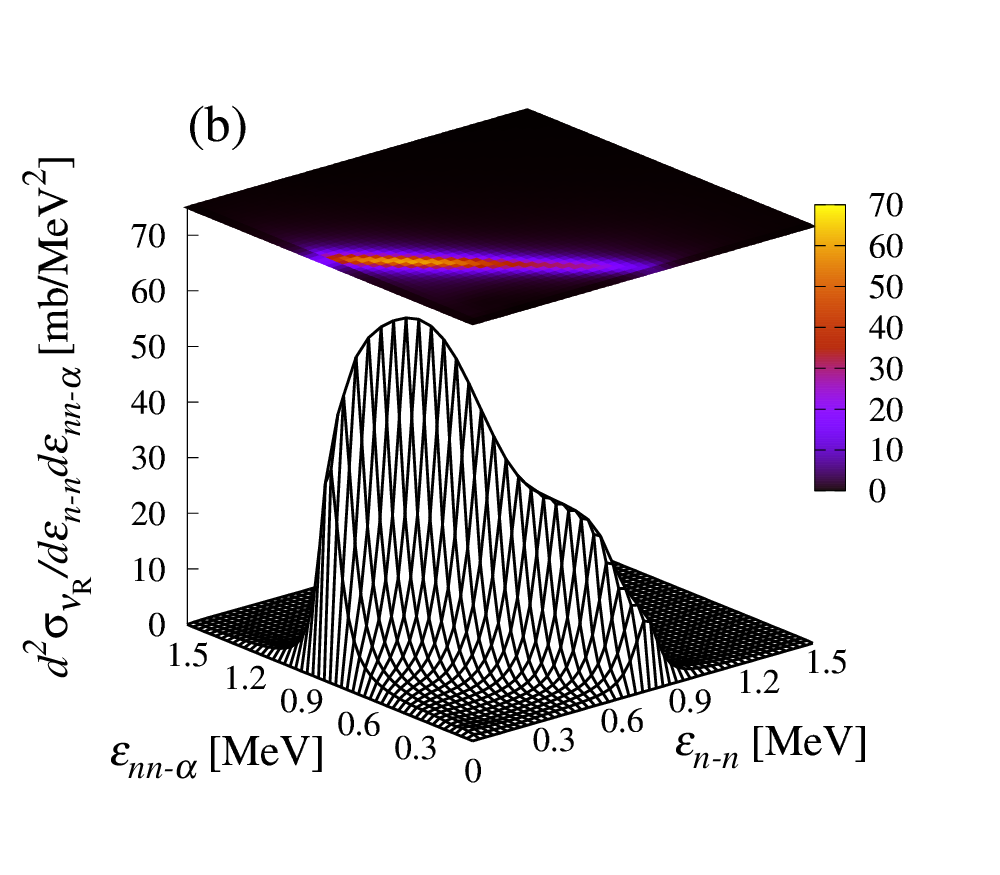}
  \caption{
  The breakup cross sections describing the transition to (a) the $2^+$ continuum states
  calculated with Eq.~\eqref{eq:DDX} and (b) the $2^+_1$ state 
  calculated with Eq.~\eqref{eq:DDX-res}.
  Here the panel (b) shows the real part of DDBUX.
  }
  \label{fig:DDBUX}
 \end{center}
\end{figure}
Next we discuss the DDBUX for the $^6$He + $^{12}$C reaction at 240 MeV/nucleon.
Figure~\ref{fig:DDBUX}(a) shows the DDBUX describing the transition to
the $2^+$ continuum states calculated with Eq.~\eqref{eq:DDX}. 
In this analysis, the OCM is not included in ${V}$ for
Eq.~\eqref{eq:csls} because we avoid the instability of numerical
results as mentioned in Ref.~\cite{Kik09}. 
The peak structure can be seen when 
$\varepsilon$ $(=\varepsilon_{n\mbox{-}n}+\varepsilon_{nn\mbox{-}\alpha})$
is around 0.8 MeV, which corresponds to the resonant energy of the $2^+_1$ state. 
This behavior is the same as shown in Fig.~1(b) of Ref.~\cite{Kik13}. 
Moreover, 
as shown in Fig.~\ref{fig:DDBUX}(b),
one clearly sees that the behavior of the DDBUX for the $2^+_1$ calculated by using 
the resonant cross section is similar to one in Fig.~\ref{fig:DDBUX}(a). 
It should be noted that the absolute value of Fig.~\ref{fig:DDBUX}(b) is larger than 
one of Fig.~\ref{fig:DDBUX}(a). 
The large absolute value can be reduced by the contributions
from the interference between the resonant state and nonresonant states
as discussed later.

\begin{figure}[tbp]
 \begin{center}
  \includegraphics[scale=0.6]{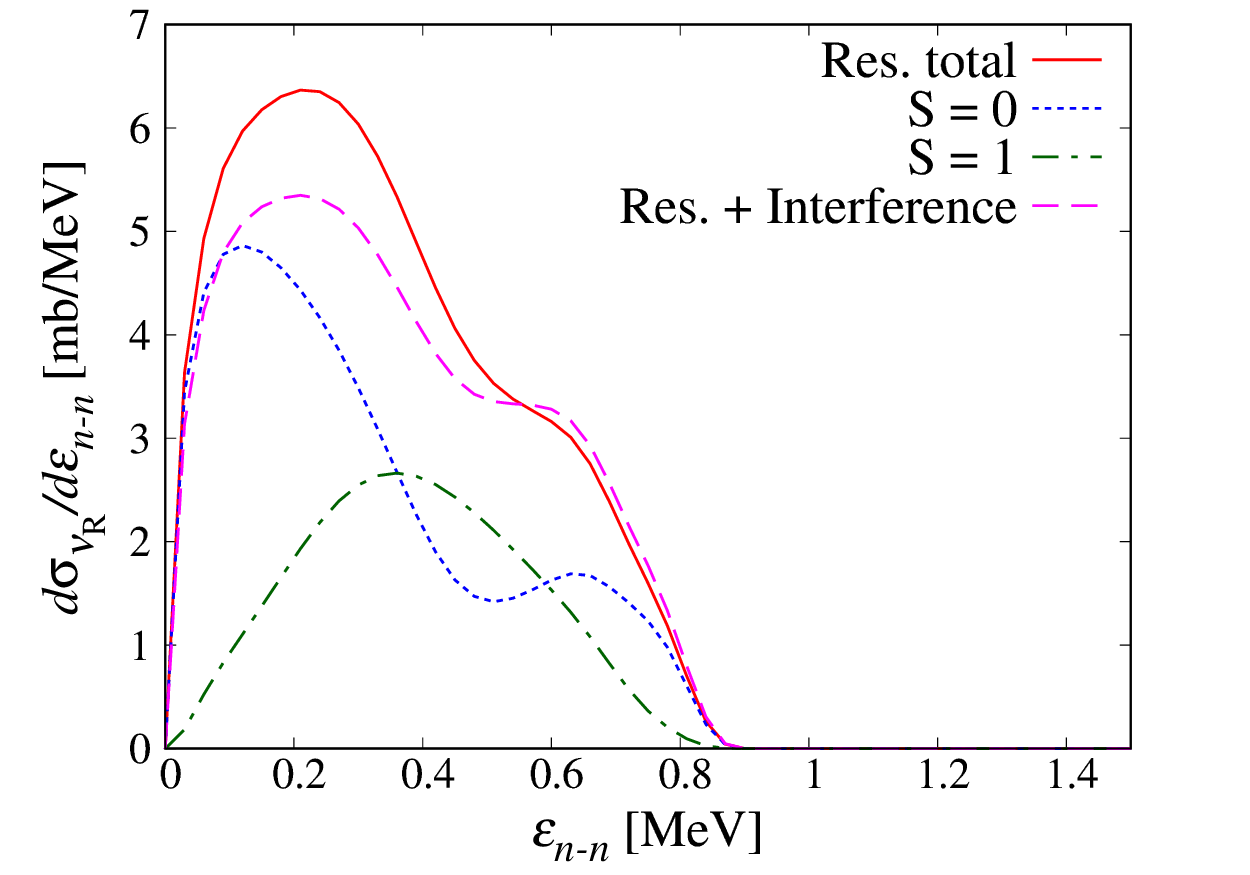}
  \caption{
  The breakup cross section with respect to $\varepsilon_{n \text{-} n}$
  calculated by using Eq.~\eqref{eq:dbux-res-nn-2}.
  }
  \label{fig:dbux-nn-res-CSM}
 \end{center}
\end{figure}
In order to investigate the dineutron in the $2^+_1$ state,
we calculate  the cross section with respect to 
the $\varepsilon_{n \text{-} n}$ as
\begin{eqnarray}
 \label{eq:dbux-res-nn-2}
  &&
  \frac{d\sigma_{\nu_{\rm R}}}{d\varepsilon_{n \text{-} n}}
  \equiv
  \int_{D}
  \frac{d^{2}\sigma_{\nu_{\rm R}}}
  {d\varepsilon_{n \text{-} n}d\varepsilon_{nn \text{-} \alpha}}
  d\varepsilon_{nn \text{-} \alpha},
  \nonumber \\
 &&~~
  (D : \varepsilon_{r} - \Gamma/2
  \le \varepsilon_{n\text{-}n} + \varepsilon_{nn\text{-}\alpha}
  \le \varepsilon_{r} + \Gamma/2) .
\end{eqnarray}
This cross section shows the energy distribution of the valence two neutrons
decaying from the resonant state. In Fig.~\ref{fig:dbux-nn-res-CSM}, 
the solid line shows the cross section, 
and the same two peaks discussed in the previous study~\cite{Kik13} are seen. 
One is the clear peak around 0.2 MeV and the other is the shoulder peak around 0.7 MeV, 
which is mentioned as the contribution from the dineutron in the $2^+_1$ state.
Because the cross section in Fig.~\ref{fig:dbux-nn-res-CSM} is reduced from only 
the $2^+_1$ state,
we can conclude that the shoulder peak confirmed in the previous study
comes from the $2^+_1$ state, not the nonresonant states.

To investigate the shoulder peak in more detail,
we separate the cross section into the $S=0$ and 1 components.
To this end,
the scattering wavefunction is represented as follow,
\begin{eqnarray}
 \label{eq:csls-spin}
  &&
 \bra{\Phi^{(-)}_{\varepsilon}(\bm{k},\bm{K})}
  =
  \bra{\Phi^{(-)}_{\varepsilon,S=0}(\bm{k},\bm{K})}
  +
  \bra{\Phi^{(-)}_{\varepsilon,S=1}(\bm{k},\bm{K})}
  ,~~~~~~
\end{eqnarray}
where $\Phi^{(-)}_{\varepsilon,S}$ ($S=0,1$) describes that
the two neutrons have the total spin $S$ in the asymptotic region.
Using Eq.~\eqref{eq:csls-spin}, 
Eq.~\eqref{eq:dbux-res-nn-2} is rewritten as
\begin{eqnarray}
 \label{eq:dbux-res-nn-2-spin}
 \frac{d\sigma_{\nu_{\rm R}}}{d\varepsilon_{n \text{-} n}}
 =
 \left(
 \frac{d\sigma_{\nu_{\rm R}}}{d\varepsilon_{n \text{-} n}}
 \right)_{S=0}
 +
 \left(
 \frac{d\sigma_{\nu_{\rm R}}}{d\varepsilon_{n \text{-} n}}
 \right)_{S=1}
 ,
\end{eqnarray}
where $(d\sigma_{\nu_{\rm R}}/d\varepsilon_{n \text{-} n})_S$ corresponds
to the cross section
obtained by replacing the
$\Phi^{(-)}_{\varepsilon}$ in Eq.~\eqref{eq:tmat-cs}  
to $\Phi^{(-)}_{\varepsilon,S}$.
The dotted and dot-dashed lines show the $S=0$ and 1 components,
respectively. One can see that the $S=0$ component has two peaks. 
The first peak around 0.2 MeV contributes
to the clear peak of total component, and the second
peak around 0.7 MeV effects on the
shoulder peak. 
For the second peak, the two-neutron pair has a relatively large momentum
that means a spatially compact pair in the coordinate space.
Consequently we can conclude that the shoulder peak indicates the existence of
the dineutron in the $2^{+}_{1}$ state.

Furthermore, to discuss the large absolute value of the resonant cross section,
we calculate the breakup cross section for the interference 
between the resonant and nonresonant states defined as
\begin{eqnarray}
 \label{eq:DDX-inter}
  &&
  \left(
  \frac{d\sigma}
  {d\varepsilon_{n \text{-} n}}
  \right)_{\rm interference}
  \equiv
  \int_{D}
  d\varepsilon_{nn \text{-} \alpha}~
  2 {\rm Re}
  \left[
  \sum_{\nu\in D'} T^{\theta\dag}_{\nu}  T^{\theta}_{\nu_{\rm R}}   
  \right.
  \nonumber \\
 &&~~\times
  \left.
 \int d\bm{k} d\bm{K} d\bm{P}~
 f^{\theta\dag}_{\varepsilon,\nu}(\bm{k},\bm{K})
 f^{\theta}_{\varepsilon,\nu_{\rm R}}(\bm{k},\bm{K})
 \delta_{\rm e.c.}
 \right],
 \nonumber \\
 &&~~
  (D' : \varepsilon_{r} - \Gamma/2
  \le {\rm Re}[\varepsilon^{\theta}_{\nu}] \le \varepsilon_{r} + \Gamma/2
  ,~~\nu \ne \nu_{R}) .
\end{eqnarray}
Here $\nu$ satisfies the region $D'$,
that is, Eq.~\eqref{eq:DDX-inter} means the interference from the nonresonant states
near the resonant energy of the $2^+_1$ state.
The dashed line means the sum of the solid line and Eq.~\eqref{eq:DDX-inter}.
Therefore the effect of the interference reduces the breakup cross section
without changing its shape.
In this analysis, we confirmed that the nonresonant contributions,
which are the terms for $\nu=\nu'\ne\nu_{\rm R}$ in Eq.~\eqref{eq:DDX-2},
and the interference between the nonresonant states are negligible.
Further the absolute value of the dashed line would be smaller when we expand the 
region $D'$.

\begin{figure}[tbp]
 \begin{center}
  \includegraphics[scale=0.6]{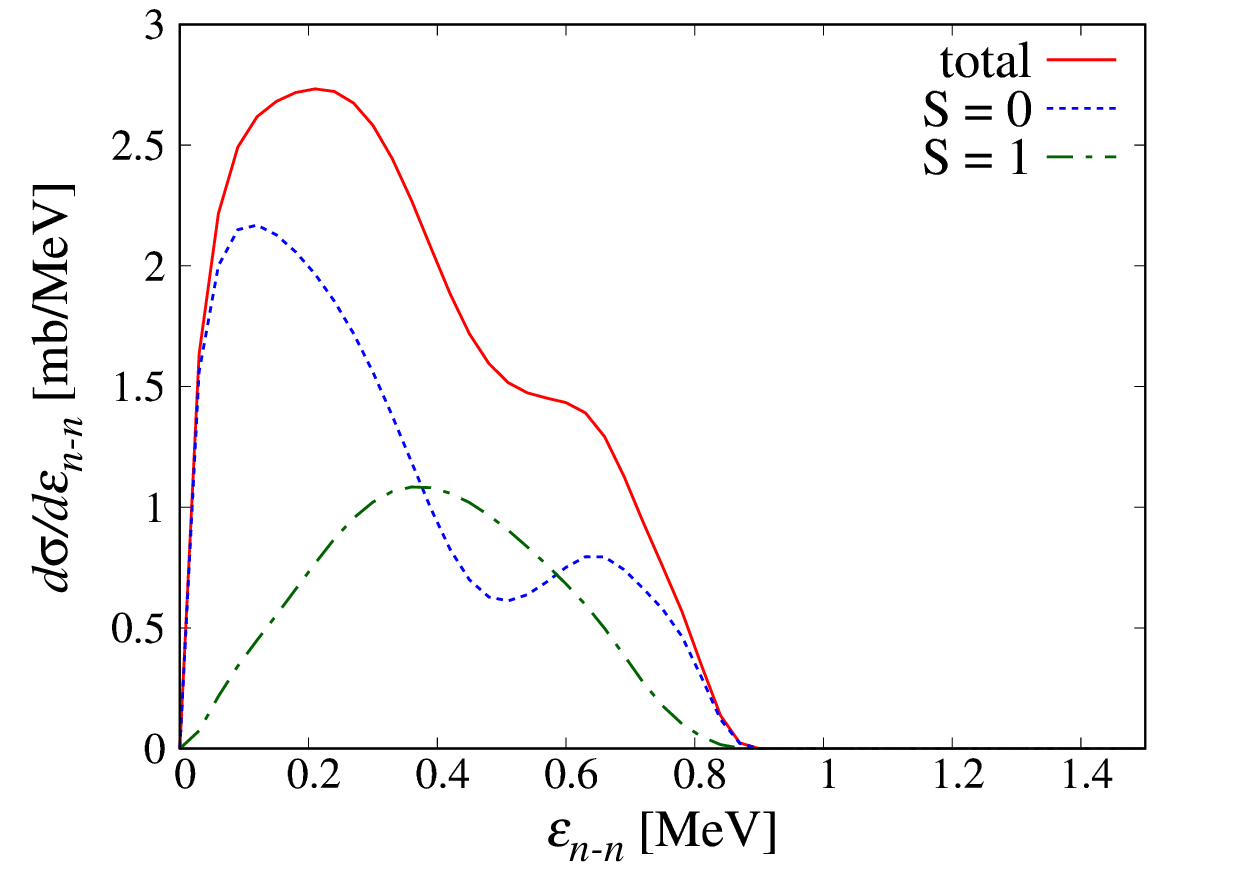}
  \caption{
  The breakup cross section with respect to $\varepsilon_{n \text{-} n}$
  calculated by using Eq.~\eqref{eq:dbux-res-nn-1}.
  }
  \label{fig:dbux-nn-res}
 \end{center}
\end{figure}
Next, to evaluate the contribution from the dineutron in the $2^+_1$ 
state on the cross section, which can be observed practically, we calculate
the cross section with respect to $\varepsilon_{n \text{-} n}$ defined
in Ref~\cite{Kik13} as 
\begin{eqnarray}
 \label{eq:dbux-res-nn-1}
  &&
  \frac{d\sigma_{2^+_1}}{d\varepsilon_{n \text{-} n}}
  \equiv
  \int_{D}
  \frac{d^{2}\sigma}
  {d\varepsilon_{n \text{-} n}d\varepsilon_{nn \text{-} \alpha}}
  d\varepsilon_{nn \text{-} \alpha}.
\end{eqnarray}
Here $d^{2}\sigma/d\varepsilon_{n \text{-} n}d\varepsilon_{nn \text{-} \alpha}$ 
is the component of the $2^+$ continuum states as 
shown in Fig.~\ref{fig:DDBUX}(a).
In Fig.~\ref{fig:dbux-nn-res}, the solid line describes 
the obtained cross section,
and the shoulder peak is also seen in the present result.
The dotted and dot-dashed lines represent the results of the $S=0$ and
1 components, respectively.
The behavior of the cross section in Fig.~\ref{fig:dbux-nn-res} is consistent 
with that in Fig.~\ref{fig:dbux-nn-res-CSM}.
Thus the cross section gated within the resonant energy region
corresponds to that for the resonant state.

\begin{figure}[tbp]
 \begin{center}
  \includegraphics[scale=0.6]{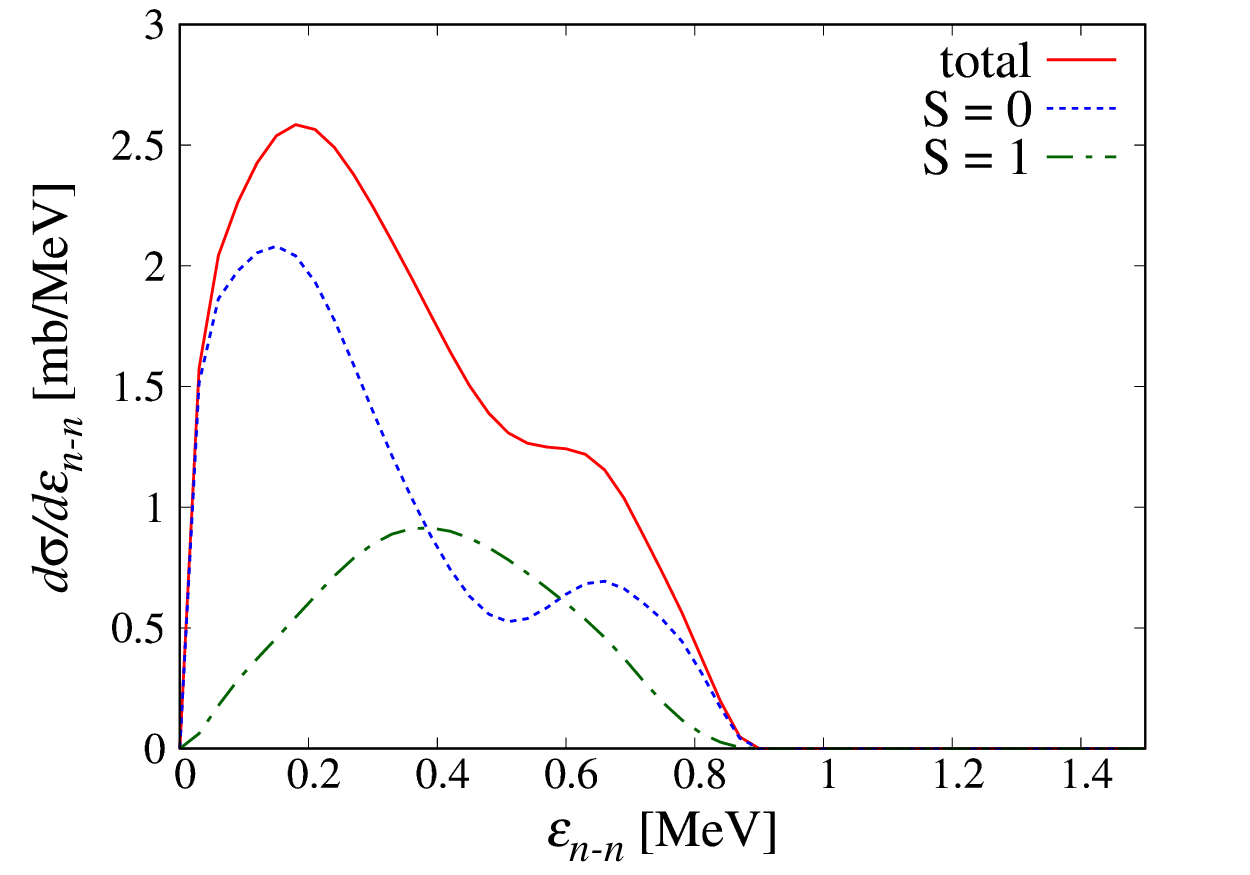}
  \caption{
  Same as Fig.~\ref{fig:dbux-nn-res}, but with
  the Gogny-Pires-Tourreil interaction as $v_{nn}$.
  }
  \label{fig:dbux-nn-res-Gogny}
 \end{center}
\end{figure}

Finally, we investigate the dependence of the dineutron structure on the interaction
between the two neutrons $v_{nn}$ in $^6$He.
As another $v_{nn}$,
we use the Gogny-Pires-Tourreil interaction~\cite{Gog70},
which has been successful in several thee-body calculations for 
core + $n$ + $n$~\cite{Zhu93,Tho04,Cas13}. 
In Fig.~\ref{fig:dbux-nn-res-Gogny},
the solid line shows the breakup cross section calculated with Eq.~\eqref{eq:dbux-res-nn-1}.
The dotted and dot-dashed lines represent the $S$ = 0 and 1 components, respectively.
One can see the same shoulder peak as one obtained with the Minnesota interaction 
as $v_{nn}$.
Thus the dineutron structure appears in the $2^+_1$ state with the reliable $v_{nn}$.
Furthermore we confirm that the optical potential
does not depend on the dineutron structure because the $T$ matrix including the effect
of the optical potential is just a constant coefficient of the resonant cross section.

%
{\it Summary.}
We analyzed the DDBUX of the $^6$He + $^{12}$C reaction at 240 MeV/nucleon
to investigate the dineutron in the resonant state $2^+_1$. 
To eliminate the nonresonant contribution from the DDBUX, 
we defined the DDBUX for the resonant state by reconstructing the 
transition matrix with the extended completeness relation in the CSM. 
The calculated cross section for the resonant state
as a function of $\varepsilon_{n\mbox{-}n}$ has the shoulder peak, 
which is discussed as the contribution from the dineutron.
Thus we found that the shoulder peak comes from the resonant state, 
not nonresonant state. 
Furthermore, we separated the cross section into the $S=0$ and 1 components.
As the result, 
the $S=0$ component of the cross section has the second peak around the shoulder peak. 
In the second peak, the two-neutron pair has a relatively large momentum 
that corresponds to a spatially compact configuration between the two neutrons. 
Therefore the shoulder peak is expected to indicate the existence of the dineutron 
in the $2^{+}_{1}$ state,
and the dineutron structure does not depend on $v_{nn}$ and the optical potential.
In the cross section, which can be observed practically,
the same peak is confirmed in the $S$ = 0 component. 
These results strongly support the suggestion in the previous study.
One of the important point of this study is that we can investigate a 
structure of a resonant state by using the resonant cross section.
In addition,
the shape of the resonant cross section does not depend on the reaction system
because the $T$ matrix is just a constant coefficient for the resonant cross section.
In the forthcoming paper,
we analyze several resonant states of other two-neutron halo nuclei,
such as $^{11}$Li, $^{14}$Be, and $^{22}$C by using the resonant cross section.

%
\section*{Acknowledgments}

The authors would like to thank Prof. Kikuchi for fruitful discussions. 
This work is supported in part by Grant-in-Aid for Scientific Research
(No.\ JP18K03650)
from Japan Society for the Promotion of Science (JSPS).


%
\end{document}